\documentclass[oneside,12pt]{article}
\usepackage{graphics,color,epsfig}
\usepackage{natbib}
\def\hbn{{\hfil\break\noindent}}

\newcommand{\ms}{\,{\rm m\,s}^{-1}}

\topmargin = -\topskip
\advance\topmargin by -\headsep
\begin{document}

\begin{center}
\Large

{\bf MARVELS: Revealing the Formation and Dynamical Evolution}
{\bf of Giant Planet Systems}
\normalsize

\bigskip

White Paper for the Astro2010 PSF Science Frontier Panel

Submitted by the SDSS-III Collaboration
\end{center}

\hbn
\underline{Contact Information:}

\hbn
Jian Ge,\\
University of Florida,\\
MARVELS Principal Investigator,\\
jge@astro.ufl.edu

\hbn
Daniel Eisenstein,\\
University of Arizona,\\
SDSS-III Director,\\
deisenstein@as.arizona.edu

\newpage
\centerline{\bf ABSTRACT}

MARVELS, the Multi-Object APO Radial Velocity Exoplanet Large-area Survey,
is a 6-year program to characterize the distribution of gas giant planets
with orbital periods ranging from several hours to two years.  MARVELS will
use multi-fiber interferometric spectrographs on the wide-field, 2.5-meter
Sloan Foundation telescope at Apache Point Observatory to monitor 
$\sim 11,000$ stars in the magnitude range $V=8-12$, visiting each star 
$\sim 30$ times over an 18-month interval, with velocity precision of 
$\approx 14$, 18, and $35\ms$ at $V=8$, 10, and 12.  MARVELS will survey
far more stars with a wider range of spectral types and metallicities than
previous radial velocity searches, yielding a statistically well defined
sample of $\sim 150$ giant planets drawn from a host sample with well 
understood selection biases.  With a unique combination of large numbers
and well characterized sensitivity, MARVELS will provide a critical data
set for testing theories of the formation and dynamical evolution of giant
planet systems.  The MARVELS detections will also be an ideal sample for
follow-up observations to identify multiple planet systems and understand
the influence of giant planet migration on the formation and survival of 
less massive planets.  MARVELS is one of four surveys that comprise 
SDSS-III (the Sloan Digital Sky Survey III), a 6-year program that will 
use highly multiplexed spectrographs on the Sloan Foundation Telescope to 
investigate cosmological parameters, the history and structure of the 
Milky Way galaxy, and the population of giant planet systems.

\smallskip
\noindent{\bf 1. Introduction}

The success of extra-solar planet searches has been one of the
most remarkable astronomical developments of the last two decades.  While
a variety of search methods are now bearing fruit, most of the known planets
have been discovered by radial velocity surveys, and these samples are the
basis of most statistical characterizations of the exoplanet population.
Exoplanets reveal an astonishing diversity of masses, semi-major axes and 
eccentricities, from the short period ``hot Jupiters", to planets in very 
elongated orbits, to planetary systems with multiple Jupiter-mass planets, 
to the super-Earth-mass planets with orbital periods of a few days (see
Udry \& Santos 2007 for a review of the properties of known exoplanets).  If any single statement 
captures the development of this field, it is that the observations have
continually revealed unanticipated diversity of planetary systems.

The standard, core accretion scenario of giant planet formation 
(Pollack et al.\ 1996) predicts that planets like Jupiter
form in nearly circular orbits, with periods of several years or more.
Growth is initiated by coalescence of icy bodies, which cannot
survive close to the parent star; once the solid core reaches
$5-10$ Earth masses its gravity is strong enough to rapidly accrete
surrounding gas and grow to Jupiter-like masses.
The two greatest surprises of extra-solar planetary discoveries
have been that many giant planets have periods below one year,
sometimes as short as one day, and that many of these planets are
on highly eccentric rather than circular orbits.
The first finding suggests that many giant planets ``migrate''
inward after their formation.  Several migration mechanisms
have been proposed, including dynamical interactions
between planets and their natal gas disks, and strong
planet-planet gravitational scattering.  The latter migration mechanism
may also explain the origin of the high orbital eccentricities.

The Multi-Object APO Radial Velocity Exoplanet
Large-area Survey (MARVELS) will use fiber-fed interferometric
spectrographs to monitor the radial velocities of 11,000 bright stars,
with the precision and cadence needed to detect giant planets with orbital
periods ranging from several hours to two years.  
Our forecasts predict that MARVELS will discover $\sim 150$ new planets,
mostly in the range of $0.5-10$ Jupiter masses.
The large sample size, comprehensive coverage of stellar hosts,
and well-defined statistical sensitivity will make MARVELS a
critical data set for testing models of the origin and dynamical
evolution of giant planet systems and the phenomenon of giant
planet migration.  MARVELS will complement other searches, which
will be sensitive to lower-mass or longer-period planets but will
not detect nearly as many `dynamically evolved' giant planets,
i.e., giant planets in the intermediate period
regime where dynamical evolution was likely important in shaping the distribution
of planet properties.

MARVELS is part of SDSS-III, a six-year program (2008-2014) that will use
existing and new instruments on the 2.5-m Sloan Foundation Telescope to
carry out four spectroscopic surveys on three scientific themes:
dark energy and cosmological parameters;
the structure, dynamics, and chemical evolution of the Milky Way;
and the architecture of planetary systems.\footnote{Because the different
SDSS-III surveys are relevant to different Astro2010 survey panels, we
are providing three separate White Papers, though the general material
on the SDSS is repeated.  A detailed description of SDSS-III is available
at {\tt http://www.sdss3.org/collaboration/description.pdf}.} All data from
SDSS-I (2000-2005) and SDSS-II (2005-2008), fully calibrated and accessible
through efficient data bases, have been made public in a series of roughly
annual data releases, and SDSS-III will continue that tradition.
SDSS data have supported fundamental work across an extraordinary range of
astronomical disciplines, including the large-scale structure of the
universe,
the evolution and clustering of quasars, gravitational lensing, the
properties
of galaxies, the members of the Local Group, the structure and stellar
populations of the Milky Way, stellar astrophysics, sub-stellar objects,
and small bodies in the solar system.  A summary of some of the major
scientific contributions of the SDSS to date appears in the Appendix.

By the time of the Astro2010 report, we hope that SDSS-III
fund-raising will be complete.  We are providing SDSS-III 
White Papers to the Astro2010 committee and panels mainly
as information about what we expect to be a major activity
for the first half of the next decade, and about data sets that
will shape the environment for other activities.
We also want to emphasize the value of supporting projects of this scale, 
which may involve public-private partnerships and international
collaborations like the SDSS, and thus the importance of 
maintaining funds and mechanisms to support the most meritorious 
such proposals that may come forward in the next decade.

\smallskip
\noindent{\bf 2. Description of MARVELS}

The key technical innovation behind MARVELS is a multi-fiber
instrument that combines a fixed-delay interferometer with moderate
dispersion spectrographs (Erskine \& Ge 2000; Ge 2002; Ge et al.\
2002).  This approach allows simultaneous, high throughput, high
velocity precision measurements of many objects using reasonable
detector sizes.  This technology is well matched to the wide field (7
deg$^2$) spectroscopic capability of the Sloan Foundation 2.5-m
telescope.  The first MARVELS instrument is a 60-fiber spectrograph
with a monolithic interferometer.  We anticipate adding a second,
similar instrument in late 2010.  MARVELS operates during bright time,
while other SDSS-III (BOSS and SEGUE-2) surveys operate in dark time;
from 2011-2014 MARVELS will carry out simultaneous operations with the
infrared Galactic structure survey APOGEE, sharing the focal plane.
The SDSS optical spectrographs will be used to obtain optical spectra
and stellar parameters of all program stars, so the properties of the
full target population will be well characterized.

The first MARVELS instrument was commissioned in September 2008.  With
this instrument, we have demonstrated calibration lamp image stability
equivalent to $\sim 3~{\rm m~s^{-1}}$  RMS stellar RV precision. On-sky, even in the
presence of higher-than-normal environmental disturbances during our
commissioning observations, our first preliminary analysis achieved
$\sim 12~{\rm m~s^{-1}}$ RMS RV precision on the RV-stable star HD
9407 ($V=6.5$) in 144 second exposures, only $\sim 30\%$ larger than
the photon-noise limit of $\sim 9~{\rm m~s^{-1}}$.  For the known
planet-bearing star TrES-2 ($V=11.4$), we achieved $\sim 30~{\rm
m~s^{-1}}$ RV precision in 40-60 minute exposures over $\sim 6$
commissioning nights. See Figure 1.  Overall, in 50-minute exposures,
the MARVELS spectrograph yields photon-noise limited radial velocity
errors of approximately 3.5, 8.5, and $21.5\ms$ at $V=8$, 10, and 12.
The minimum goal for velocity precision including all sources of
systematic error is 14, 18, and $35\ms$ at these magnitudes, though we
expect improvements in the analysis pipeline and modest improvements
in the instrument thermal control to bring the measurement precision
closer to the photon-noise limit.  Normal MARVELS survey operations
commenced in Oct.\ 2008, and we have already detected several
candidate radial velocity variables.  

Detailed simulations of observing strategies prior to the survey
commencement led to a design in which each program star is visited $\sim 30$ times over
an 18-month period, with sampling that provides sensitivity on all
period scales within this interval.  Assuming that a second 60-fiber
instrument is available for the final four years of the survey,
MARVELS will search about 10,000 main sequence
stars and 1,000 giant stars for planets.  Figure~2 shows the
projected sensitivity of MARVELS in different mass and period ranges,
based on (conservative) estimates of radial velocity errors consistent with 
the initial survey data and given above.  Forecasts based on currently known planet statistics
predict that MARVELS will detect $\sim 150$ planets in the mass
range $M\sin i \approx 0.5 - 10 M_{\rm Jup}$.  In the mass and period
range that MARVELS probes, it will provide a factor of several 
increase over the largest homogeneous and statistically complete
samples currently available (Fischer \& Valenti 2005; Cumming et al.\ 2008).

\begin{figure}[t]
\begin{center}
\includegraphics[angle=90,width=3.2in]{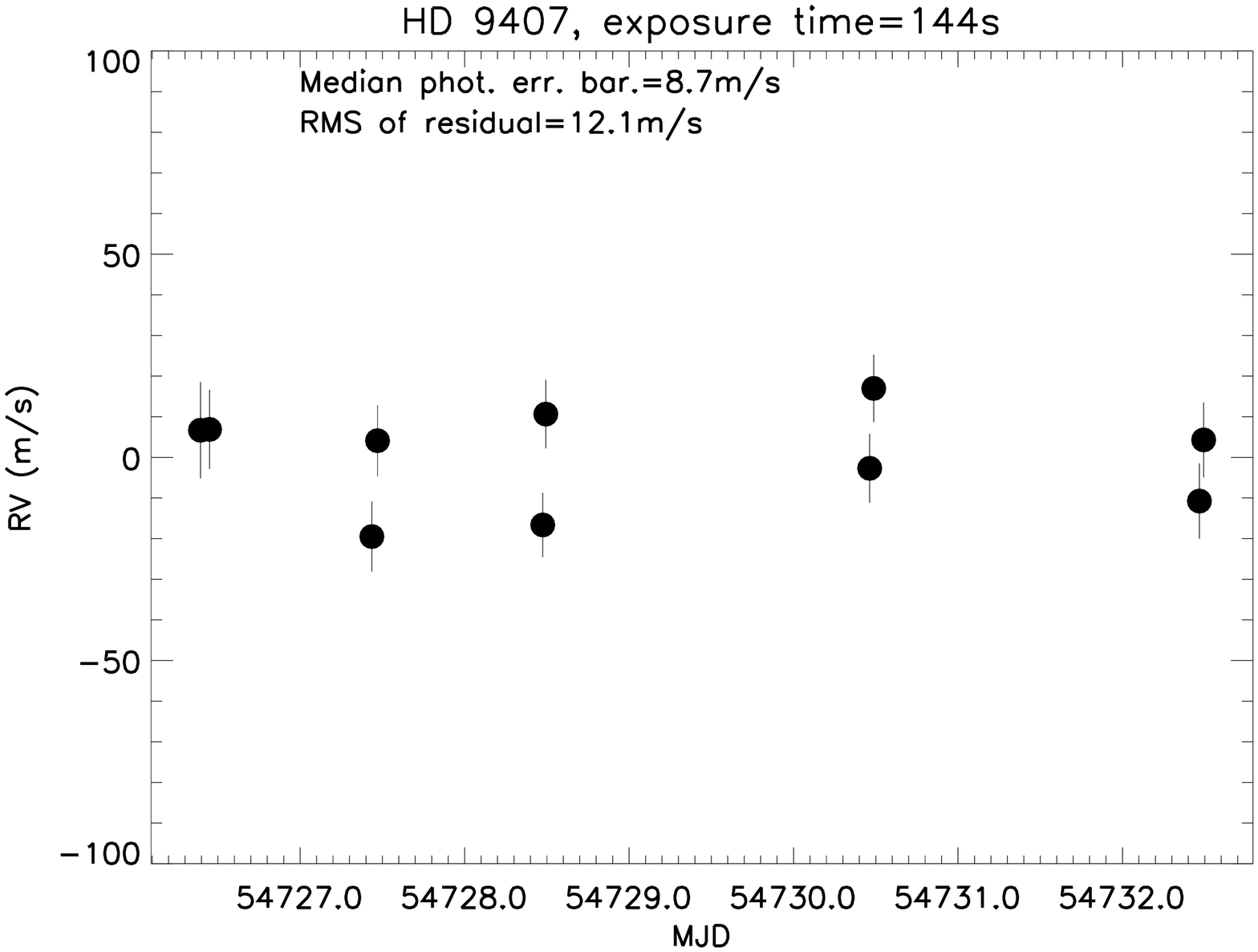}
\includegraphics[angle=90,width=3.2in]{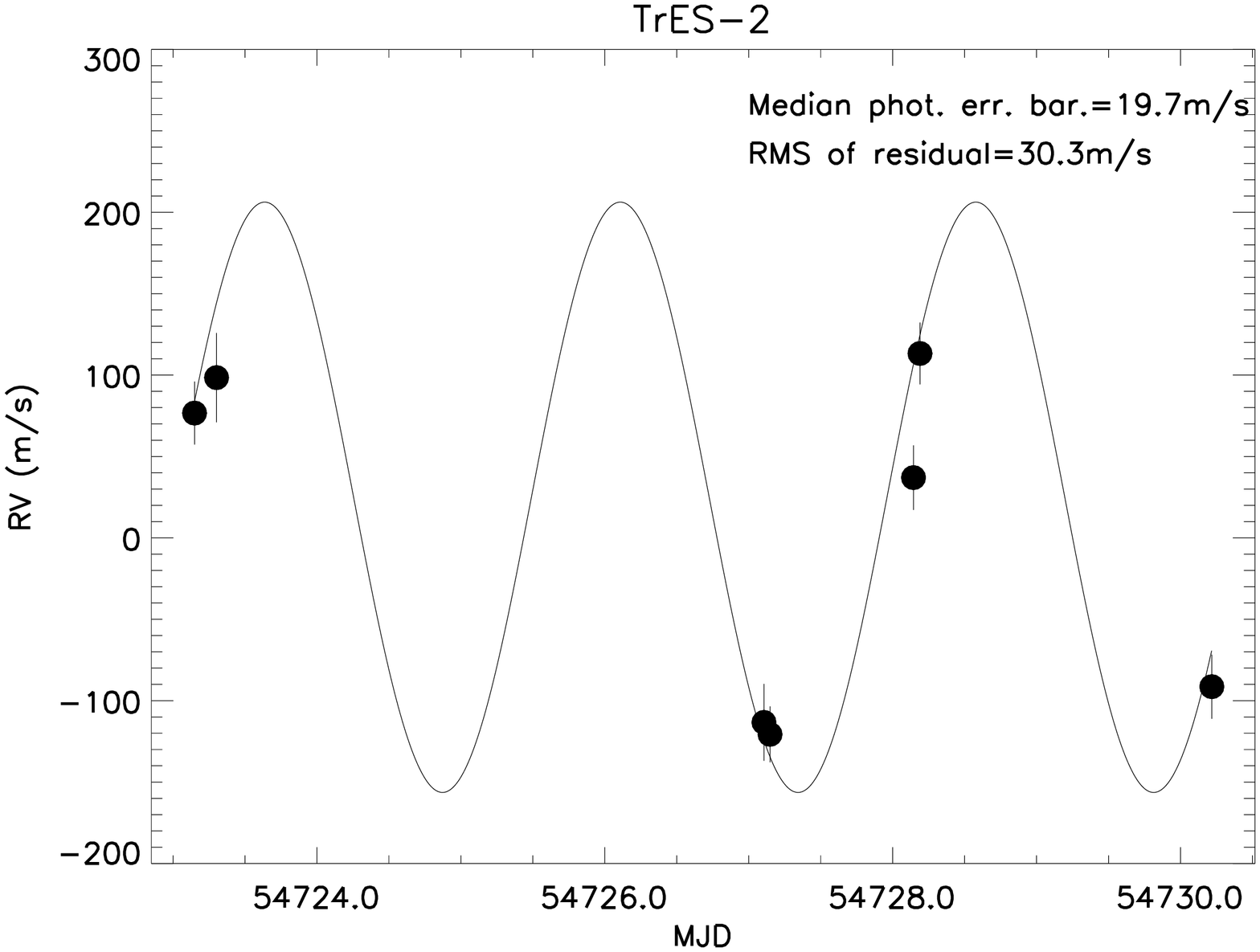}
\end{center}
\vskip-0.2in
\caption{
MARVELS Sep.\ 2008 commissioning observations of HD 9407 (left),
a known RV-stable star with $V=6.6$, and TrES-2, a known planet-hosting
star with $V=11.4$.}
\label{fig:commdata}
\end{figure}

\begin{figure}[t]
\begin{center}
\includegraphics[width=6.0in]{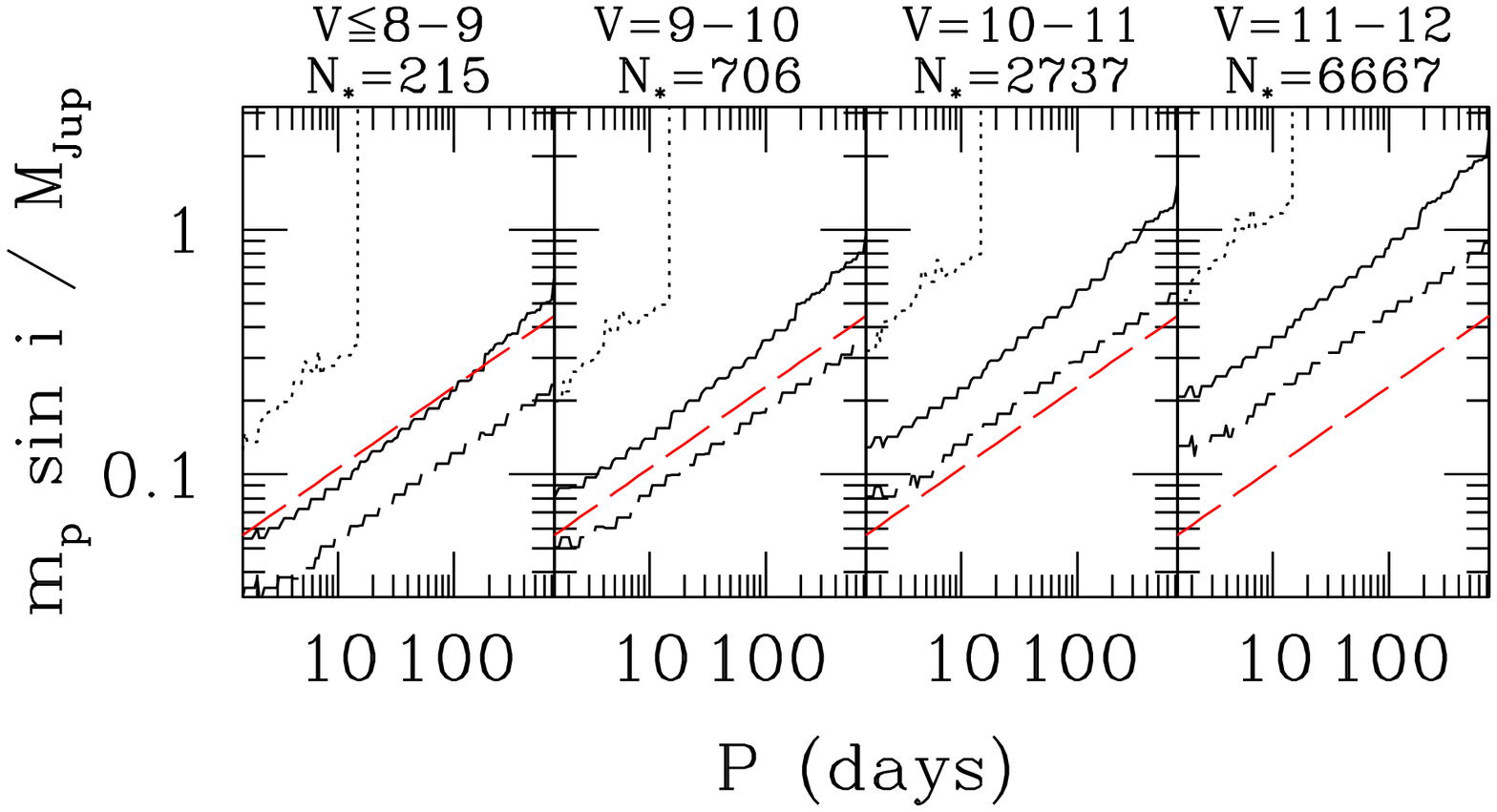}
\end{center}
\vskip-0.2in
\caption{
The efficiency of MARVELS planet detection for four different
ranges of stellar apparent magnitude $V$.
The number of stars in each magnitude range that would be monitored
in the baseline survey is listed above each panel.
Lines show contours of constant planet detection efficiency for
host stars of the indicated brightness.
At each period $P$, MARVELS would detect 95\% of planets
above the mass threshold $m_p\sin i$ indicated by the dotted line,
50\% above the solid line, and 5\% above the dashed line.
Long-dashed lines show the locus of a planet with a 10 m/s
orbital velocity in a circular orbit about a solar-mass star.
}

\label{fig:marvelsefficiency}
\end{figure}

MARVELS will provide a powerful data set for testing the emerging,
detailed models of planet formation, migration, and dynamical evolution
(e.g., Ida \& Lin 2004; Alibert et al.\ 2005; Kornet et al. 2005, Ford \& Rasio 2008).  
These models incorporate different assumptions about initial conditions and
about the physical mechanisms that govern planet growth and orbital evolution.
They make quantitative statistical predictions for the joint distributions
of planet mass, semi-major axis, and orbital eccentricity, and for the
dependence of planet frequency and orbital properties on the mass and
metallicity of the host star.  The predictions can be tested by
combining them with star-by-star efficiency calculations,
and these tests can incorporate
both detections and non-detections in a statistically rigorous fashion.
The short- and intermediate-period systems
probed by MARVELS are especially important for understanding the physics
of planet migration and for testing theories that address the
puzzling preponderance of planets with large eccentricities.

The broad selection of target stars will make MARVELS ideal for
studying the correlation of planetary systems with stellar metallicity,
mass, multiplicity, age, evolutionary stage, activity level, and
rotation velocity.  
For example, the observed correlation of planet frequency with metallicity 
(Santos et al.\ 2004; Fischer \& Valenti 2005) may favor 
the core accretion model of planet formation over the gravitational instability model (Boss 1997),
or it may simply be a signature of a metallicity-dependent migration rate (Livio \& Pringle 2003).
By targeting a  large number of lower metallicity stars,
MARVELS will provide a more robust determination of the planet-metallicity correlation,
and so provide important constraints on models that seek to explain it.  

The host stars of the MARVELS planet detections will be
valuable targets for longer term, higher precision radial velocity
monitoring to detect longer period and/or lower mass companions.
For example, we would like to know what
fraction of systems with migrated giant planets have surviving 
lower mass planets at smaller separations, or massive long-period
planets.  With a well characterized primary survey and a well designed
follow-up program, the two-stage selection effects of such
a survey are straightforward to apply to predictive models.
The large number of targets also makes MARVELS sensitive to rare
classes of planetary systems that would be largely missed by smaller
surveys: systems in the ``brown dwarf desert,'' massive hot Jupiters, rapidly
interacting multiple planet systems, very-hot Jupiters, and/or
planets with extremely high eccentricities.  

Finally, a wide range of auxiliary science can also be extracted from
the MARVELS dataset.  For example, MARVELS will discover a large
number of binary stars, which can be used to construct a relatively
unbiased sample of binary orbital elements, as well as determine the
close binary fraction as a function of primary mass and metallicity.
Some fraction of these binary systems will also be eclipsing, and
follow-up RV and photometric observations of particularly interesting
eclipsing binary (EB) systems can be used to test models of low-mass
stars and brown dwarfs. Double-lined eclipsing systems can be used to
test stellar models by providing 
model-independent physical properties (i.e.\ masses, radii,
gravities) for both stars, and eclipsing systems with very low-mass 
secondaries can be used to study the mass-radius
relationship of objects near or below the hydrogen burning limit.

\smallskip
\noindent{\bf 3. Conclusions}

The great scientific opportunities in the field of extra-solar planets
have sparked an explosion of new instruments and new methods for
planet searches.  In terms of the range of planet masses and
separations to which it is sensitive, MARVELS will complement ongoing
and planned radial velocity, transit, microlensing, and astrometric
surveys, which probe lower mass or longer period systems but will not
yield a comparably large sample of `dynamically evolved' giant planets.
The regime probed by MARVELS is crucial to understanding the physics
of giant planet migration and dynamical interaction, and thus important
for understanding the overall physics of the formation and
evolution of planetary systems.  MARVELS will also set the stage for
larger-scale, higher-precision multi-fiber radial velocity searches,
which will likely be pursued in the second half of the decade. 
These future surveys will provide
exoplanet samples vastly larger than those that exist today, with
radial velocity precisions of $\sim 1~{\rm m~s^{-1}}$, similar
to the best precisions available with current single-object RV instruments,
and sufficient to detect planets with mass as low as several times the 
mass of the Earth. 

In round numbers, SDSS-III is a \$40 million project, and the
funding is largely in hand thanks to generous support from
the Alfred P. Sloan Foundation, the National Science Foundation,
the Department of Energy, and the Participating Institutions
(including international institutions and participation groups
supported, in some cases, by their own national funding agencies).
The SDSS has demonstrated the great value of homogeneous
surveys that provide large, well defined, well calibrated data
sets to the astronomical community.  In many cases, such surveys
are made possible by novel instrumentation, and they often
require multi-institutional teams to carry them out.
The case for supporting ambitious surveys in the next decade is
best made by considering
the contributions of the SDSS to the astronomical breakthroughs
of the {\it current} decade, as summarized in the Appendix below.

\smallskip
\noindent{\bf Appendix: The SDSS Legacy}

The SDSS (York et al.\ 2000) is one of the most ambitious and influential
surveys in the history of astronomy.
SDSS-II itself comprised three surveys: the Sloan Legacy Survey completed
the goals of SDSS-I, with imaging of 8,400 square degrees and spectra of
930,000 galaxies and 120,000 quasars; the Sloan Extension for Galactic
Understanding and Exploration (SEGUE) obtained 3500 square degrees of
additional imaging and spectra of 240,000 stars; and the Sloan Supernova
Survey
carried out repeat imaging ($\sim 80$ epochs) of a 300-square degree area,
discovering nearly 500 spectroscopically confirmed Type Ia supernovae for
measuring the cosmic expansion history at redshifts $0.1 < z < 0.4$.
Based on an analysis of highly cited papers, Madrid \& Machetto
(2006, 2009) rated the SDSS as the highest impact astronomical
observatory in 2003, 2004, and 2006 (the latest year analyzed so far).
The final data release from SDSS-II was made public in October, 2008,
so most analyses of the final data sets are yet to come.

The list of extraordinary scientific contributions of the SDSS
includes, in approximately chronological order:
\begin{itemize}
\setlength{\itemsep}{-4pt}
\setlength{\parsep}{0pt}
\item{} {\it The discovery of the most distant quasars,}
tracing the growth of the first supermassive black holes and
probing the epoch of reionization.
\item{} {\it The discovery of large populations of L and T dwarfs,}
providing, together with 2MASS, the main data samples for systematic
study of sub-stellar objects.
\item{} {\it Mapping extended mass distributions around galaxies with weak
gravitational lensing,} demonstrating that dark matter halos extend to
several hundred kpc and join smoothly onto the larger scale dark matter
distribution.
\item{} {\it Systematic characterization of the galaxy population,}
transforming the study of galaxy properties and the correlations among them
into a precise statistical science, yielding powerful insights
into the physical processes that govern galaxy formation.
\item{} {\it The demonstration of ubiquitous substructure in the outer Milky
Way,} present in both kinematics and chemical
compositions, probable signatures of hierarchical buildup of
the stellar halo from smaller components.
\item{} {\it Demonstration of the common origin of dynamical asteroid
families,} with distinctive colors indicating similar
composition and space weathering.
\item{} {\it Precision measurement of the luminosity distribution of
quasars,} mapping the rise and fall of quasars and the growth of
the supermassive black holes that power them.
\item{} {\it Precision measurements of large scale galaxy clustering,}
leading to powerful constraints on the matter and energy contents of the
Universe and on the nature and origin of the primordial fluctuations
that seeded the growth of cosmic structure.
\item{} {\it Precision measurement of early structure with the
Lyman-$\alpha$ forest,} yielding precise constraints on the
clustering of the underlying dark matter distribution
$1.5-3$ Gyr after the big bang.
\item{} {\it Detailed characterization of small and intermediate scale
clustering of galaxies} for classes defined by luminosity, color, and
morphology, allowing strong tests of galaxy formation theories
and statistical determination of the relation between galaxies and dark
matter halos. \item{} {\it Discovery of many new companions of the Milky Way
and Andromeda,}
exceeding the number found in the previous 70 years, and providing
critical new data for understanding galaxy formation in low mass halos.
\item{} {\it Discovery of stars escaping the Galaxy,} ejected by
gravitational interactions with the central black hole, providing
information on the conditions at
the Galactic Center and on the shape, mass, and total extent of
the Galaxy's dark matter halo. \item{} {\it Discovery of acoustic
oscillation signatures in the clustering of galaxies,} the first
clear detection of a long-predicted cosmological signal,
opening the door to a new method of cosmological measurement that
is the key to the BOSS survey of SDSS-III.
\item{} {\it Measurements of the clustering of quasars over a wide range
of cosmic time,} providing critical constraints on the dark matter
halos that host active black holes of different luminosities at
different epochs.
\end{itemize}

Half of these achievements were among
the original ``design goals'' of the SDSS, but the other half
were either entirely unanticipated or not
expected to be nearly as exciting or powerful as they turned out to be.
The SDSS and SDSS-II have enabled systematic investigation and
``discovery'' science in nearly equal measure, and we expect that tradition
to continue with SDSS-III.

Funding for SDSS-III has been provided by the Alfred P. Sloan Foundation,
the Participating Institutions, the National Science Foundation, and the
U.S. Department of Energy. The SDSS-III web site is http://www.sdss3.org/.

SDSS-III is managed by the Astrophysical Research Consortium for the
Participating Institutions. The SDSS-III Collaboration is still growing; at
present, the Participating Institutions are the University of Arizona, the
Brazilian Participation Group, University of Cambridge, University of
Florida, the French Participation Group, the German Participation Group, the
Joint Institute for Nuclear Astrophysics (JINA), Johns Hopkins University,
Lawrence Berkeley National Laboratory, Max Planck Institute for Astrophysics
(MPA), New Mexico State University, New York University, Ohio State
University University of Portsmouth, Princeton University, University of
Tokyo, University of Utah, Vanderbilt University, University of Virginia,
and the University of Washington.

\hbn{\bf References}\\
\smallskip
Alibert, Y., Mordasini, C., Benz, W., \& Winisdoerffer, C.\ 2005, A\&A, 434, 343\\
Boss, A.~P.\ 1997, Science, 276, 1836\\
Cumming, A., et al.\ 2008, PASP, 120, 531\\
Erskine D., \& Ge, J. 2000, in ASP Conference Series 195, 501\\
Fischer, D.A. \& Valenti, J. 2005, ApJ, 622, 1102\\
Ford, E.~B., \& Rasio, F.~A.\ 2008, ApJ, 686, 621\\
Ge, J., 2002, ApJ, 571, L165\\
Ge, J. et al. 2002, PASP, 114, 1016\\
Ida, S., \& Lin, D.~N.~C.\ 2004a, ApJ, 604, 388\\
Kornet, K., Bodenheimer, P., R{\'o}{\.z}yczka, M., \& Stepinski, T.~F.\ 2005, A\&A, 430, 1133\\
Livio, M., \& Pringle, J.~E.\ 2003, MNRAS, 346, L42\\
Madrid, J. P. \& Macchetto, F. D. 2006, BAAS 38, 1286\\
Madrid, J. P. \& Macchetto, F. D. 2009, BAAS, in press [arXiv:0901.4552]\\
Pollack, J.~B., et al.\ 1996, Icarus,  124, 62\\
Santos, N.C., Israelian, G., \& Mayor, M. 2004, A\&A 415, 1153\\
Udry, S., \& Santos, N.C. 2007, ARA\&A, 45, 397\\
York D.G., et al., 2000, AJ, 120, 1579\\


\end{document}